\documentclass[prl,aps,twocolumn,amsfonts,amssymb,graphicx]{revtex4}

\begin{document}

\title{Distillation protocols: Output entanglement and local mutual information}

\author{
Micha{\l} Horodecki\(^1\), Jonathan Oppenheim\(^{1,2}\), 
Aditi Sen(De)\(^3\), and Ujjwal Sen\(^3\)
}


\affiliation{$^{1}$Institute of Theoretical Physics and Astrophysics,
University of Gda\'nsk, 80-952 Gda\'nsk, Poland}
\affiliation{$^{2}$Department of Applied Mathematics and Theoretical Physics, University of
Cambridge, Cambridge, U.K.}
\affiliation{\(^{3}\)Institut f\"ur Theoretische Physik, Universit\"at Hannover, D-30167 Hannover,
Germany}

\begin{abstract}

A complementary behavior between local mutual information and average output entanglement 
is derived for arbitrary bipartite ensembles. This leads to bounds on the yield 
of entanglement in  distillation protocols that involve disinguishing. This bound 
is saturated in the hashing protocol for distillation, for Bell-diagonal states.  
\end{abstract}

\maketitle

\newtheorem{lemma}{Lemma}
\newtheorem{corollary}{Corollary}
\newtheorem{theorem}{Theorem}

\font\Bbb =msbm10
\font\eufm =eufm10
\def\Real{{\hbox{\Bbb R}}} \def\C{{\hbox {\Bbb C}}}
\def\spec{R_{\alpha\beta}}
\def\id {{\hbox{\Bbb I}}}
\def\kvec{{\bf k}}
\def\lvec{{\bf l}}
\def\duzomniejsze{<\kern-.7mm<}
\def\duzowieksze{>\kern-.7mm>}
\def\intlarge{\mathop{\int}\limits}
\def\textbf#1{{\bf #1}}
\def\be{\begin{equation}}
\def\ee{\end{equation}}
\def\ben{\begin{eqnarray}}
\def\een{\end{eqnarray}}
\def\eea{\end{array}}
\def\bea{\begin{array}}
\newcommand{\bei}{\begin{itemize}}
\newcommand{\eei}{\end{itemize}}
\newcommand{\bee}{\begin{enumerate}}
\newcommand{\eee}{\end{enumerate}}

\def\ra{\rangle}
\def\la{\langle}
\def\blacksquare{\vrule height 4pt width 3pt depth2pt}
\def\lcal{{\cal L}}
\def\dcal{{\cal D}}
\def\pcal{{\cal P}}
\def\hcal{{\cal H}}
\def\trace{\mbox{Tr}}
\def\dt#1{{{\kern -.0mm\rm d}}#1\,}

\def\tr{{\rm Tr}}
\def\id{{\rm I}}
\def\ra{\rangle}
\def\la{\langle}
\def\>{\rangle}
\def\<{\langle}
\def\blacksquare{\vrule height 4pt width 3pt depth2pt}
\def\ic{I_{coh}}
\def\ot{\otimes}
\def\rhoab{\varrho_{AB}}
\def\sigmaab{\sigma_{AB}}
\def\rhoa{\varrho_{A}}
\def\sigmaa{\sigma_{A}}
\def\rhob{\varrho_{B}}
\def\sigmab{\sigma_{B}}
\def\PD{private distribution }

\def\pab{p_{a,b \ldots (n)}}

\emph{Introduction.}--
Distillation of entanglement 
\cite{IBMdistillation,IBMhuge} is a key issue in attaining 
nonclassical tasks in quantum communication protocols \cite{NC}. 
In a typical communication protocol, 
entanglement must be shared between distant partners (Alice and Bob). Since 
channels are invariably noisy, the partners usually end up with mixed state entanglement,
which must then be distilled into pure form via local operations and classical communication (LOCC),
to make them amenable to the envisaged quantum communication protocol.

The aim of this paper is two-fold. We obtain an upper bound on local mutual 
information, \(I^{LOCC}\), of arbitrary bipartite ensembles. We then use this bound
to provide bounds on the yield of entanglement in any distillation protocol, that 
use local distinguishing of ensembles of states. 
%
The obtained bounds are then compared with the yield in the existing distillation 
protocols (e.g. \cite{IBMdistillation,
IBMhuge, Werner-Wolf-hashing}) and similar generalizations thereof,
and also in some other cases, in which the distillation is based on 
a distinguishability protocol \cite{Walgate-twoent,ebar-thekey}. 
As a spin-off, we obtain a complementarity relation between 
local mutual information and average output entanglement.

\emph{Generalized universal Holevo-like upper bound on local mutual information.}--
To begin, we obtain a generalized Holevo-like bound on local mutual information for 
arbitrary bipartite ensembles.
Suppose then that a source prepares the ensemble \({\cal R} = \{p_x, \varrho_x^{AB}\}\) and sends the 
\(A\) part to Alice and the \(B\) part to Bob. The task of Alice and Bob is to estimate the identity \(x\)
of the sent state. If Alice and Bob are together, so that they are allowed to perform global operations,
the mutual information is bounded by the Holevo quantity \cite{Holevo}, 
\(\chi_{\cal R} = 
S(\varrho)- \sum_x p_x S(\varrho_x)\),
 where 
\(\varrho\) is the average ensemble state \(\sum_x p_x \varrho_x\).
\(S(\cdot)\) is the von Neumann entropy and is defined for a state \(\varrho\) as 
\(S(\varrho) = -\mbox{tr} \varrho \log_2 \varrho\).
We will however 
need the following result \cite{Schumacher,iacc}, which is a generalization
of the Holevo bound on mutual information.
\begin{lemma}
\label{important}
If a measurement on ensemble $Q=\left\{ p_{x},\varrho_{x}\right\}$
produces result $y$ with probability \(p_y\), and leaves a post-measurement ensemble
$Q^{y}=\left\{ p_{x|y},\varrho_{x|y}\right\}  $, 
then the mutual information \(I\) (between the identity of state in the 
ensemble and measurement outcome) extracted from the
measurement has the following bound:
\begin{equation}
I\leq\chi_{Q}-\overline{\chi}_{{Q}^{y}}.
\label{prothhom}
\end{equation}
Here $\overline{\chi}_{{Q}^{y}}$ is the average Holevo bound for
the possible post-measurement ensembles, i.e. \(\sum_y p_y \chi_{{Q}^{y}}\).
\end{lemma}

Suppose now that Alice and Bob are far apart, so that they are able to perform 
only local operations and communicate classically between the operations. 
In this scenario, universal Holevo-like upper bound on local mutual information for 
an arbitrary bipartite ensemble \(\{p_x,\varrho_x^{AB}\}\) was obtained in \cite{iacc}:
\begin{eqnarray}
\label{sei}
I^{LOCC} \leq 
S(\varrho^A) + S(\varrho^B) - \max_{Z=A,B}\sum_x p_x S(\varrho^Z_x).
\end{eqnarray}
Here \(\varrho_x^{A(B)} = \mbox{tr}_{B(A)} (\varrho_x^{AB})\), and 
\(\varrho^{A(B)} = \mbox{tr}_{B(A)}\sum_xp_x\varrho_x^{AB}\).
In this paper, we will prove a generalization of this bound. Precisely, we show that
\begin{eqnarray}
\label{asol}
I^{LOCC} 
 \leq S(\varrho^A) + S(\varrho^B) - \sum_x p_x S(\varrho^B_x)  \nonumber \\
 - \sum_{a,b, \ldots, (n)} \pab
S\left(\sum_x p_{x|ab\ldots(n)} \varrho^A_{x|ab \ldots (n)}\right).  
\end{eqnarray}
Here \(\{p_{x|ab\ldots(n)}, \varrho^{AB}_{x|ab \ldots (n)}\}\) is the post-measurement 
ensemble obtained after the measurement in the \(n\)th step, and 
\(\pab\) 
is
the probability of the 
sequence of 
measurement outcomes in steps 1, 2, \(\ldots\), \(n\). 
Our generalization in  (\ref{asol}) 
is related to the previous bound in  (\ref{sei}), in a similar way as
Lemma \ref{important} is related to the original Holevo bound.

We will now prove the inequality in (\ref{asol}). To start the protocol for obtaining the identity \(x\) 
of the given ensemble \({\cal R} = \{p_x, \varrho_x^{AB}\}\), 
Alice makes a measurement \cite{first}, and suppose that she obtains an
outcome \(a\), with probability \(p_a\). Suppose that the post-measurement
ensemble (for outcome \(a\) at Alice) is \({\cal R}_a = \{p_{x|a}, \varrho_{x|a}^{AB}\}\).



The results presented in this 
paper are in terms of mutual information, 
which when maximized over all measurement strategies
gives the ``accessible information''.
All the results are of course true for the extreme case of the best measurement strategy (for attaining 
maximal mutual information), but are  true also for any other nonextreme measurement 
strategy. The mutual information gathered from the measurement of Alice 
has the following bound due to Lemma \ref{important}:
\(
I_1^A \leq \chi_{{\cal R}^A} - \overline{\chi}_{{\cal R}^A_a}
\).
Here \(\chi_{{\cal R}^A}\) is the Holevo quantity of the \(A\) part of 
the ensemble \({\cal R}\), i.e. of the ensemble \({\cal R}^A = \{p_x, \varrho_x^A\}\).
And \(\chi_{{\cal R}^A_a}\) is the Holevo quantity of the \(A\) part of 
the ensemble \({\cal R}_a\).
The subscript \(1\) in 
\(I_1^A\) indicates that the information is extracted from the first measurement.

After Alice communicates her result to Bob, his ensemble is
 \({\cal R}_a^B = \{p_{x|a}, \varrho_{x|a}^{B}\}\), with 
\(\varrho_x^B = \mbox{tr}_A (\varrho_x^{AB})\).
Suppose now that Bob performs a measurement and obtains outcome \(b\) with probability \(p_b\),
so that the post-measurement ensemble (at his part) is \({\cal R}^B_{ab} = \{p_{x|ab}, \varrho^B_{x|ab}\}\), where
\(\varrho^B_{x|ab} = \mbox{tr}_A\left(\varrho^{AB}_{x|ab}\right)\). 
So (again due to Lemma \ref{important}), the information extracted in Bob's measurement has the 
following bound:
\(
I_2^B \leq 
\overline{\chi}_{{\cal R}^B_a} - \overline{\chi}_{{\cal R}^B_{ab}}
\).

This procedure of measuring and communicating the result goes on for an arbitrary number of steps, and by the 
chain rule for mutual information (see e.g. \cite{CoverThomas}), the mutual information obtained in 
all steps is 
\(
I^{LOCC} = I^A_1 + I^B_2 + I^A_3 + \ldots
\).
Note that this quantity depends on the measurement strategy followed by  Alice and Bob.

Now we (repeatedly) use the following facts:  

(i) The von Neumann entropy is  concave (i.e. 
\(S(p_1\varrho_1 + p_2 \varrho_2) \geq p_1S(\varrho_1) + p_2 S(\varrho_2)\), for 
arbitrary density matrices \(\varrho_1\) and \(\varrho_2\), and probabilities \(p_1\) and \(p_2\)), and positive. 

(ii)
A measurement on one subsystem cannot change the state at a distant subsystem.

(iii)
The average change (initial minus final) of von Neumann entropy due to 
a measurement on one subsystem cannot be less than the average change in a distant subsystem. So for 
example, after the first measurement by Alice, we have 
\(\sum_xp_xS(\varrho_x^A) - \sum_a p_a \sum_x p_{x|a} S(\varrho_{x|a}^A) 
\geq \sum_xp_xS(\varrho_x^B) - \sum_a p_a \sum_x p_{x|a} S(\varrho_{x|a}^B)\). 

(iv)
The Holevo quantity is positive.

Then  after \(n\) steps of measurements, we obtain the inequality (\ref{asol}).

We have assumed that the last measurement is performed by Alice. 
The last term of the bound (\ref{asol}) is a contribution from this last measurement by Alice.
We will see below that the final result is free from 
this asymmetry. Moreover, for the same measurements, but using the above items (i)-(iv) in a 
different way, one can reach the inequality (\ref{asol}), but with \(A\) and \(B\) interchanged, i.e.,
we also have
\begin{eqnarray}
\label{asol25}
I^{LOCC} 
 \leq S(\varrho^A) + S(\varrho^B) - \sum_x p_x S(\varrho^A_x)  \nonumber \\
 - \sum_{a,b, \ldots, (n-1)} p_{ab\ldots (n-1)}
S\left(\sum_x p_{x|ab\ldots(n-1)} \varrho^B_{x|ab \ldots (n-1)}\right).  
\end{eqnarray}
Note that now the last term is a contribution from the next to 
last measurement, which (due to the assumption that Alice performed the last measurement) is performed by Bob.
Inequalities (\ref{asol}) and (\ref{asol25}) give us upper bounds on local mutual information, 
for \emph{arbitrary} bipartite ensembles. 
These inequalities are true for any measurement strategy of Alice and Bob. In particular, 
they are  true for the one which maximizes \(I^{LOCC}\). This is then the so-called locally accessible 
information  (\(I_{acc}^{LOCC}\)).

The last terms in the bounds on local mutual information in inequalities (\ref{asol}) and 
(\ref{asol25}) respectively are negative quantities, 
due to the positivity of von Neumann entropy. Leaving it out, we have 
the inequality (\ref{sei}).

\emph{Input and output entanglements.}--
We now try to write the bounds on local mutual information in (\ref{asol}) and (\ref{asol25})
in a more revealing form.
To that end, note that the von Neumann entropy of either of the the local density matrices of a bipartite state 
is no smaller than the  
entanglement of formation \cite{IBMhuge}, and the entanglement of formation is a lower bound for any asymptotically 
consistent measure of bipartite entanglement \cite{HHHDonald}. 

Then, the last term in the upper bound of Eq. (\ref{asol25}) is 
\(
\leq
- \sum\limits_{a,b, \ldots, (n-1)} p_{ab\ldots (n-1)}
E\left(\sum_x p_{x|ab \ldots (n-1)} \varrho_{x|ab \ldots (n-1)}^{AB}\right)
\), 
which in turn 
(by the fact that entanglement cannot increase (on average) under LOCC) 
 is  no greater than 
\begin{eqnarray}
\label{char}
 - \sum_{a,b,\ldots,(n)} 
\pab
E\left(\sum_x p_{x|ab\ldots (n)} 
\varrho_{x|ab\ldots (n)}^{AB}\right),
\end{eqnarray}
where \(E\) denotes any asymptotically consistent measure of bipartite entanglement.
The last term of (\ref{asol}) is directly \(\leq\) the right-hand-side (rhs) of (\ref{char}), by the 
fact that the von Neumann entropy of local density matrix is \(\geq\) any asymptotic  entanglement measure.  
The rhs of (\ref{char}) (without the minus sign) is just the average entanglement that we obtain at the output 
in the \(n\) step local measurement protocol between Alice and Bob. We denote it by \(\overline{E}_{out}\).
Note that from here on, the results are independent of whether it was Alice or Bob who ended 
the protocol.

Refering back to the inequalities (\ref{asol}) and (\ref{asol25}), we have 
\begin{eqnarray}
\label{asol1}
I^{LOCC} \leq S(\varrho^A) + S(\varrho^B) - \max_{Z=A,B}\sum_x p_x S(\varrho^Z_x) 
- \overline{E}_{out}. \nonumber \\
\end{eqnarray}

It is possible to write Eq. (\ref{asol}) in an even more revealing way. Note that 
\(S(\varrho^A) + S(\varrho^B) \leq N\), where \(N\) is the number of qubits (two-dimensional 
quantum systems) in the Alice-Bob system. I.e. \(N = \log_2 d_A d_B\), where 
\(d_A\) and \(d_B\) are respectively the dimensions of the Hilbert spaces of Alice's and Bob's particles.
Moreover, we have \(S(\varrho^B_x) \geq {\cal E}(\varrho_x^{AB})\), 
where 
again \({\cal E}\) denotes any asymptotically consistent measure of bipartite entanglement 
\cite{IBMhuge, HHHDonald}. 
The quantity \(\sum_x p_x {\cal E}(\varrho_x^{AB})\) is the average input (initial) entanglement 
in the Alice-Bob system. We denote it by \(\overline{{\cal E}}_{in}\). 
We use a separate notation for the asymptotic entanglement measure for the input states than that 
in the output states, to underline the fact that they can be different measures. It is known 
that there exist several asymptotically consistent measures of bipartite entanglement (see \cite{michalQIC}).
We will come back to this point later. 
So finally we have 
\begin{eqnarray}
\label{ghyama}
I^{LOCC} \leq N - \overline{{\cal E}}_{in}
- \overline{E}_{out}.
\end{eqnarray}
Eq. (\ref{ghyama}) can also be obtained from Eq. (\ref{asol25}), with the additional assumption of 
monotonicity under LOCC of \(E\).
Before connecting above bounds on local mutual information with 
entanglement distilled in distillation protocols, let us note some interesting features of these
inequalities.

\emph{Complementarity between extracted and unused information.}-- 
One way of interpreting the result in Eq. (\ref{ghyama}) is to note that 
the terms \(I^{LOCC}\) and \(\overline{E}_{out}\) depend on the measurement 
protocol followed by Alice and Bob. The other two terms (\(N\) and \(\overline{{\cal E}}_{in}\))
are fixed for a given ensemble. So, writing the inequality as 
\(I^{LOCC} + \overline{E}_{out}
\leq N - \overline{{\cal E}}_{in}\),
we see that the left hand side can be interpreted as a sum of ``extracted information'' (\(I^{LOCC}\))
and ``unused information'' (\(\overline{E}_{out}\)). Independently (i.e. considered 
separately), the extracted and unused 
informations depend on the measurement strategy followed by Alice and Bob. However for all 
strategies, the sum of the extracted and unused informations is bounded by \(N - \overline{{\cal E}}_{in}\).  

\emph{On bound entanglement with nonpositive 
partial transpose.}--
Another interesting feature of the inequality (\ref{ghyama}) is that 
the entanglement measures \(E\) and \({\cal E}\) need not be the same measures. 
They must only be be no greater 
than the von Neumann entropy of either of the local density matrices.
In particular, any asymptotically consistent measure of bipartite entanglement satisfy such 
conditions (see \cite{michalQIC}). 
This may have nontrivial consequences. 
For example, we may require that 
\({\cal E}\) must be a convex function, and keep \(E\) to be such that it need not necessarily be  convex
\cite{convex}. 
The only entanglement measure for which there is some evidence for nonconvexity  is for 
distillable entanglement \cite{IBMhuge}, and this is related to 
the phenomenon of bound entanglement \cite{HHHbound}. Precisely, it was shown in Ref. \cite{Shor} 
that distillable entanglement can be proven to be nonconvex, if there exists a certain 
bound entangled state \cite{NPTbound}, having nonpositive partial transpose (NPT) \cite{PPT}. 
Bound entanglement, and more particularly NPT bound entanglement 
is not a well understood phenomenon of quantum mechanics. 
We believe that the inequality (\ref{ghyama}), 
may have important consequences for NPT bound entangled states. 
The point that we make here is also to be seen with respect to the fact that, below we actually 
relate the output entanglement \(E_{out}\) to entanglement distilled in different distillation
protocols, and bound entanglement is precisely that entanglement which cannot be distilled.

\emph{Bound on entanglement distillable via protocols correcting all errors.}--
We will now consider distillation protocols based on full
distinction between the possible pure states in a decomposition 
of \(m\) copies a bipartite state \(\rho\). 
Suppose therefore that Alice and Bob share \(m\) copies of the state $\rho$ given by 
\be
\rho=\sum_i p_i |\psi_i\>\<\psi_i|.
\ee
where $|\psi_i\>$ are eigenvectors of \(\rho\). 
Alice and Bob can imagine that they actually share some string 
of the form $\psi_{i_1}\ot \ldots \ot\psi_{i_m}$.
Now we propose the following strategy for distillation.
Alice and Bob try to fully distinguish between all strings.
I.e. they apply some LOCC operation, that tells them 
what is the string that they share. Usually during such distinguishing, 
they  destroy  the string  to some degree. 
For example, the protocol of distinguishing two pure orthogonal 
states given in \cite{Walgate-twoent}, destroys the states completely.
Yet in the hashing protocol for distilling entanglement, Alice and Bob are able to distinguish 
strings without destroying all entanglement they share \cite{IBMhuge}.

In the case of full distinguishing (in some distillation protocol \(P\)), 
the accessible information is $mS(\rho)$. The initial entanglement per input pair
is equal to $\overline S_A\equiv \sum_ip_i S(\rho^A_i)$,
where $\rho_i^A$ is 
the local density matrix of 
$|\psi_i\>$.
Since we have full distinguishing, the final entanglement 
is pure entanglement, so that it can be converted reversibly by LOCC, into singlets \(|\psi^-\>
= \frac{1}{\sqrt{2}}(|01\> - |10\>)\) \cite{BBPS1996}.
Thus the output entanglement is the entanglement \(D_P\) that has been distilled 
in such protocol $P$.
Using the inequality (\ref{asol1}) we have then 
\be
S  \leq  S_A +S_B - \overline S_A - D_P,  
\ee
where for ease of notation, we have used the notations \(S \equiv S(\rho)\), 
\(S_A \equiv S(\tr_B \rho)\),
 and 
\(S_B \equiv S(\tr_A \rho)\). 
This gives 
\be
\label{Sagor29}
D_P\leq S_A + S_B - S -\overline S_A.
\ee
Note that since \(|\psi_i\>\) are pure,
 \(\overline S_A = \sum_i p_i S(\tr_B|\psi_i\>\<\psi_i|) 
= \sum_i p_i S(\tr_A|\psi_i\>\<\psi_i|) = \overline S_B \). So the last 
term in the above inequality (\ref{Sagor29}) can be replaced by \(\overline S_B \).
For the case of Bell diagonal states (i.e. states that are 
diagonal in the canonical maximally entangled basis 
\cite{Bell-diagonal}), we have $S_A= S_B =\overline S_A = \log_2 d$ 
so that in that case, inequality (\ref{Sagor29}) gives us
\be
D_P (\rho) \leq \log_2  d -S(\rho).
\ee
This result is compatible with the fact that the quantity $\log_2 d-S(\rho)$ 
can be attained by hashing methods that reveal all errors 
\cite{IBMhuge,Werner-Wolf-hashing}. 

It is also instructive to consider a hypothetical protocol,
in which Alice and Bob would divide their \(m\) systems into two groups 
$G_1$ and $G_2$ of length 
$m_1$ and $m - m_1$ respectively. 
Now by applying some LOCC actions, 
Alice and Bob would aim to get to know the identities of the states of systems 
from $G_1$, while $G_2$ would serve as a resource to do this
and would be destroyed during protocol. The protocol differs from 
the previous one, as in the present case, Alice and Bob does not aim to distinguish 
between states of systems from this latter group. 

Suppose now that such a protocol (\(P'\)) exists. 
Then the output entanglement  is $m_1 \overline S_A$, the input one is $m \overline S_A$, while 
the mutual information is equal to $m_1S(\rho)$. 
The entanglement $D_{P'}$ distillable in this 
protocol is therefore equal to the 
output entanglement divided by $m$:
 \[D_{P'}= \frac{m_1 \overline S_A}{m}.\]
We obtain the following constraint for $r\equiv {m_1\over m}$:
\be
r\leq {S_A+S_B - \overline S_A \over S + \overline S_A}
\ee
which finally leads to
\be
D_{P'}\leq {S_A+S_B - \overline S_A \over S + \overline S_A} \overline S_A.
\ee
(We remember that \(\overline S_A = \overline S_B\).)
For Bell diagonal states 
it gives the following bound:
\be
\label{Mana-di}
D_{P'}(\rho)\leq {(\log_2d)^2 \over \log_2d + S(\rho)}
\ee
(For Bell diagonal states in \(2 \otimes 2\), this reduces to 
\(D_{P'}(\rho)\leq {1 \over 1 + S(\rho)}
\).)
The bound is always nonzero, even for separable states.
This means that the inequality (\ref{asol1}) is not the only restriction 
on local mutual information in this complicated situation. This is however not surprising, as
in the considered protocol, we assumed that using a part of the string, 
we can get the whole information about the rest of the string,
but nothing about the used part.
What one expects is that at the some point, one perhaps would also 
gain some information about the used part.  Note here that the bound in (\ref{Mana-di})
is for those distillation protocols in which one bases on a distinguishing protocol.

\emph{Conclusions.}-- We have shown that it is possible to obtain  bounds on 
the yield in 
distillation protocols, basing on distinguishability, of bipartite states,
from a complementarity connecting local 
mutual information with average output entanglement, for the case of bipartite 
ensembles. For Bell-diagonal states, saturation of this bound is obtained 
in the hashing protocol for distillation. It is consistent with 
results of \cite{shor-smolin}, where to beat hashing bound, degenerate codes 
were applied. Whether any distillation protocol is a
distinguishing process remains an open question.

\emph{Note added.}-- After completion of our work, we came 
to know of the recent related work in Ref. \cite{India}.


MH is supported by the Polish Ministry of Scientific
Research and Information Technology under the (solicited) grant No.
PBZ-MIN-008/P03/2003 and by EC grants RESQ and QUPRODIS.
%
%
JO is supported by  EC grant PROSECCO. 
%
AS and US acknowledge  support  from  the Alexander von Humboldt Foundation.


\begin{thebibliography}{10}
\bibitem{ebar-thekey} Henceforth, by ``distinguishing'', we will mean ``local distinguishing''.
\bibitem{first} The results obtained are independent of whether it was Alice or Bob who started the protocol.
\bibitem{convex} An entanglement measure \(E\) is said to be convex, if for all bipartite 
density matrices \(\varrho_1\) and \(\varrho_2\), and probabilities \(p_1\) and \(p_2\), 
we have 
\(
E(p_1 \varrho_1 + p_2 \varrho_2)  \leq  p_1 E(\varrho_1) + p_2 E(\varrho_2). 
\)
This inequality essentially says that forgetting of information is not useful 
to increase entanglement. 
If this inequality does not hold for some choice of the density matrices (and probabilities),
then the corresponding entanglement measure is said to be nonconvex.
\bibitem{Bell-diagonal} The canonical 
maximally 
entangled states in \(d \otimes d\) 
are
\(
\frac{1}{\sqrt{d}}\sum_{j=0}^{d-1} e^{2\pi ijn/d}\left|
j\right\rangle \left|  (j+m)\mbox{mod}\quad d\right\rangle\), 
(\(n,m=0,\ldots , d-1\)). 
For \(d = 2\),
it
is the familiar Bell basis.
\end{thebibliography}
\end{document}